\RequirePackage{fix-cm}
\documentclass{article}                     
\usepackage{graphicx}
\usepackage{graphicx}
\usepackage{amsmath}
\usepackage{amsfonts}
\usepackage{dirtytalk}
\usepackage{csvsimple}
\usepackage[table]{xcolor}
\usepackage{enumitem}
\usepackage{cite}
\usepackage{adjustbox}
\usepackage{float}
\usepackage{apacite}
\usepackage{placeins}
\usepackage{lineno}
\usepackage{algorithm2e}
\usepackage{url}
\usepackage{xcolor}
\usepackage{setspace}
\usepackage{rotating}
\usepackage{caption}
\usepackage{textcomp}
\usepackage{cite}
\usepackage{amssymb}
\usepackage{multicol}
\usepackage{tabularx}
\usepackage[flushleft]{threeparttable}
\usepackage[margin=1.3in]{geometry}
\usepackage{apacite}
\usepackage[labelsep=endash]{caption}
\usepackage{ragged2e}
\usepackage{extdash}
\RestyleAlgo{ruled} 
\usepackage{authblk}

\date{}
\title{The R.O.A.D. to clinical trial emulation}

\author[1]{Dimitris Bertsimas}
\author[1]{Angelos Koulouras}
\author[2]{Hiroshi Nagata}
\author[1]{Carol Gao}
\author[2]{Junki Mizusawa}
\author[2]{Yukihide Kanemitsu}
\author[1,3]{Georgios Antonios Margonis}
\affil [1]{Sloan School of Management and Operations Research Center, E62-560, Massachusetts Institute of Technology, Cambridge, MA, 02139, USA.}
\affil[2]{National Cancer Center Hospital, Tokyo, Japan.}
\affil[3]{Department of Surgery, Memorial Sloan Kettering Cancer Center, New York, NY, 10065, USA.}

\begin{document}
\maketitle
\noindent *Georgios Antonios Margonis, MD, PhD; (Corresponding Author) email: margonig@mskcc.org\\

\begin{abstract}
\noindent
 
Observational studies provide the only evidence on the effectiveness of interventions when randomized controlled trials (RCTs) - apart from the initial RCT that establishes the efficacy of a treatment compared to a placebo - are impractical due to cost, ethical concerns, or time constraints. While many methodologies aim to draw causal inferences from observational data, there is a growing trend to model observational study designs after hypothetical or existing RCTs, a strategy known as \say{target trial emulation.} Despite its potential, causal inference through target trial emulation is challenging because it cannot fully address the confounding bias inherent in real-world data due to the lack of randomization. In this work, we present a novel framework for target trial emulation that aims to overcome several key limitations, including confounding bias. The framework proceeds as follows: First, we apply the eligibility criteria of a specific trial to an observational cohort derived from real-world data. We then \say{correct} this cohort by extracting a subset that, through optimization techniques, matches both the distribution of covariates and baseline prognoses (i.e., the prognosis in the trial's control group) of the target RCT. Next, we address unmeasured confounding by adjusting the prognosis estimates of the treated group to align with those observed in the trial, using cost-sensitive counterfactual models. Following trial emulation, we go a step further by leveraging the emulated cohort to train optimal decision trees, developed by our team, to identify subgroups of patients exhibiting heterogeneity in treatment effects (HTE). The absence of confounding is verified using two external models, and the validity of the treatment effects estimated by our framework is independently confirmed by the team responsible for the original trial we emulate. To our knowledge, this is the first framework to successfully address both observed and unobserved confounding, a challenge that has historically limited the use of randomized trial emulation and causal inference in general since the 1950s. Additionally, our framework holds promise in advancing precision or personalized medicine by identifying patient subgroups that benefit most from specific treatments.

\end{abstract}

\newpage

\section{Introduction}

Since the 1950s, observational data analyses have been widely used to approximate the goals of randomized clinical trials (RCTs), serving as a cornerstone in comparative studies across health, medicine, and related fields \cite{cochran1972statistical, dorn1953philosophy, robins1986new, rubin1974estimating, wold1954causality}. In 2019, the European Medicines Agency acknowledged the need to complement randomized controlled trials with alternative methods, particularly for research questions where traditional trials are unfeasible or unethical. Similarly, the 21st Century Cures Act in the United States mandated that the Food and Drug Administration (FDA) provide guidance on how real-world data can support drug approvals \cite{ramagopalan2020can, eichler2020novel, cave2019real}. In the UK, the National Institute for Health and Care  Excellence (NICE) released in 2022 its \say{Real-World Evidence Framework,} emphasizing the importance of using a target trial framework for estimating treatment effects from observational data to inform regulatory decision-making \cite{NICE2022}.

This growing endorsement by policymakers of the use of real-world data, traditionally reserved for RCTs, to inform clinical decision making has prompted researchers to try to emulate hypothetical or existing RCTs to solve clinical problems \cite{admon2019emulating}. For example, according to a recent systematic review, only between 2020 and 2022, 168 studies explicitly designed to emulate a target trial were published across 26 fields of medicine \cite{hansford2023reporting}. However, these efforts have faced significant challenges, particularly due to the inability to control for unobserved or unmeasured confounding factors. This issue manifests itself in clinical data in two primary ways. First, although sicker patients are more likely to receive treatment, while lower-risk patients are typically assigned to observation, some of the prognostic factors guiding these decisions may not be reported in clinical datasets. Second, even if we could fully account for the known prognostic factors that influence treatment decisions, there are often additional biological factors - unknown but coexisting with the known ones - that contribute to outcomes in sicker patients. This problem of incomplete knowledge is pervasive in medicine, as evidenced by the low to moderate C-indices of many prognostic models. Unlike in RCTs, where randomization balances both known and unknown factors between treatment groups, this remains an unsolvable issue in real-world data. In both cases, the result is that the baseline prognosis of treated patients is typically worse than that of untreated patients, leading to biased comparisons of treatment efficacy. This can result in underestimating the true benefit of treatment or, in some cases, suggesting that treatment might harm patients. These biases complicate the accurate assessment of treatment effectiveness using real-world data. 

Although various methods such as matching, e.g., on the propensity score (PS), or its machine learning applications, such as gradient-boosted PS models and multi-layer perceptron neural network-based PS models), stratification, regression, standardization, inverse probability weighting (IPW), g-estimation, and doubly robust methods - can address observed confounding, these approaches are ineffective against unobserved (unmeasured) confounding, as it cannot be directly accounted for \cite{zang2023high, hernan2022target}. Even Hernán and Robins, who pioneered a widely adopted methodology for emulating RCTs, acknowledge unobserved confounding as a limitation \cite{hernan2016using}. Specifically, they state: \say{Despite correctly emulating all other components of a target trial, observational analyses may be invalidated if confounding cannot be adequately adjusted for. Sophisticated adjustment methods are sometimes necessary, but they work only when good data on confounders are available.} \cite{hernan2021methods}. 

Our recent work introduced the R.O.A.D. framework, which offers clinically interpretable treatment recommendations from observational data by counteracting the effects of unobserved confounding \cite{bertsimas2024road}. In the first step, the framework matches treated and untreated patients with similar estimated baseline prognoses, improving comparability and mimicking the design of an RCT. Despite controlling for baseline risk, unobserved confounders may still influence results. To address this, the second step of the R.O.A.D. framework incorporates a weighting system that places greater emphasis on patients who received treatment and had favorable outcomes, compensating for biases related to baseline disease severity. This approach enhances the accuracy of prognosis estimates and enables the construction of optimal decision trees, which provide actionable treatment recommendations based on patient characteristics. Unlike traditional RCTs, optimal decision trees not only offer insights into the average treatment effect but also reveal how different subgroups of patients respond to treatment, thus providing more tailored guidance for clinical decision-making \cite{amram2022optimal}.

While the two-step approach introduced in R.O.A.D. is versatile, it is particularly effective for treatments with high efficacy, where assigning larger weight to patients who received treatment and had positive outcomes aligns with the treatment's effectiveness. However, determining how much to adjust the estimated treatment effect using this method is unclear for a broader range of treatments, such as those with low or moderate efficacy, unless the target trial is specified. In this work, we modify the prognostic matching of the R.O.A.D. framework to obtain a matched cohort with both a similar covariate distribution and baseline prognosis to the target RCT. The counterfactual estimation is also corrected by a weighting process, which ensures that the estimated treatment effects in each treatment group align with those reported in the target trial. Specifically, we \say{correct} both the cohort, ensuring its covariates' distribution and baseline prognoses match those of a target RCT, and the prognosis estimates of the treated group, thus addressing unmeasured confounding by aligning them with those observed in a specific trial. To ensure the robustness of this approach, the new framework is externally validated in the cohort of the target RCT, guaranteeing a fair and accurate validation process. Of note, this is the first work that transforms observational data into a target RCT cohort and provides treatment recommendations with external validation in the target RCT, without having access to the target RCT cohort.

\section{Methods}\label{sec1}

In this section, we establish the theoretical foundation of our framework by using an example. We focus on an observational dataset involving patients with colorectal liver metastases (CRLMs) who are undergoing post-surgery follow-up and are eligible to receive either chemotherapy or observation. Our objective is to emulate the JCOG0603 trial, a phase II/III RCT that compared the outcomes of hepatectomy with chemotherapy (mFOLFOX) versus hepatectomy alone in 300 patients with liver-only metastases \cite{kanemitsu2021hepatectomy}. The disease-free survival (DFS) endpoint used in the RCT is particularly relevant for our analysis, as it allows us to avoid potential immortal bias. The trial's DFS endpoint includes recurrence, death, and the development of secondary cancers. In our observational dataset, we record recurrence and death, but not secondary cancers, which means our endpoint is recurrence-free survival (RFS). This distinction is primarily semantic, as the difference between DFS and RFS is minimal - only 13 of the 176 events in the trial were secondary cancers. This is, in fact, a strength rather than a limitation, as successfully validating our framework using RFS instead of DFS for model training would demonstrate the robustness of our approach and further reinforce our findings. Moreover, this does not impact the external validation of our framework, as our collaborators, who designed JCOG0603, used RFS (which was not reported in the trial) for external validation. In the remainder of this section, we outline the methods used to transform the observational dataset into a trial-like dataset and describe how we provide optimal treatment recommendations for patient subgroups, defined by combinations of patient characteristics.





\subsection{Inclusion Criteria}

First, we apply the eligibility criteria of the target trial to our observational cohort. Given that drug licensing systems internationally effectively require RCT evidence of benefit or equivalence for nearly all new drugs, it is almost always possible to find an existing trial to emulate \cite{paez2022beyond}. The key inclusion criteria for the trial used to demonstrate our novel framework were: patient age of 20-75 years, histologically confirmed liver metastasis from colorectal cancer, R0 resection, liver metastasis as the first and only recurrence in metachronous cases, no extrahepatic metastasis or recurrence, and no other chemotherapy or radiotherapy within 3 months prior to enrollment. We queried the international, multi-institutional cohort of centers participating in the International Genetic Consortium for Colorectal Liver Metastasis (IGCLM) to identify patients who met the target trial's eligibility criteria \cite{margonis2018association}.

\subsection{Risk Stratification and Prognostic Matching}

We extend the prognostic matching methodologies from the R.O.A.D. framework to achieve two objectives: (a) to create a matched cohort in which each treatment group has the same baseline risk or prognosis (i.e., the risk of the outcome in the absence of treatment), and aligns with baseline prognosis of the control group in the trial, and (b) to match the distribution of covariates between the observational data and the trial data. Note that the first step reduces the confounding inherent in observational data. For example, the JCOG0603 trial reported a 5-year DFS rate of 38.7\% for the control group, and approximately 79\% of trial participants had left-sided tumors in each treatment group. Our matching algorithm selects patients who both have similar baseline recurrence risks across different treatment groups (mimicking the random assignment of treatment in a randomized controlled trial) and forms a cohort with a baseline 5-year RFS of 38.7\%, with 79\% having left-sided tumors. We describe each step in more detail below.

Let us assume that we have an observational dataset of $n$ patients with CRLM. Each patient $i$ is characterized by a vector of covariates $\boldsymbol{x}_{i}$, including features such as disease-free interval (DFI), tumor site, and tumor size. Let $t$ denote the treatment in general, and $t_{i}$ denote the treatment of patient $i$, where $t_{i}=1$ if the patient received chemotherapy after hepatectomy, and $t_{i}=0$ otherwise. The outcome $y$ is defined as $y_{i}=1$ if the patient experienced recurrence, and $y_{i}=0$ otherwise. Thus, the characteristics of patient $i$ are summarized by $(\boldsymbol{x}_{i}, t_{i}, y_{i})$ for $i=1, \cdots, n$. Note that in survival models, we are also interested in time-to-event variables (e.g., time from resection to recurrence). However, for simplicity, we limit the discussion to predicting the event itself.

The first step is to stratify patients based on the distribution of the risk of recurrence, assuming no treatment was provided. To estimate the baseline risk distribution of the patients, we train a machine learning (ML) model, $g_{\theta}$, with parameters $\theta$, using the untreated patients $i \in \mathcal{S}_{0}$ to predict the outcome based on their covariates. Specifically, $w_{i} = g_{\theta}(\boldsymbol{x}_{i})$ represents the probability that $y_{i} = 1$, or the probability that patient $i$ experienced a recurrence. This model can then be applied to each treated patient $i \in \mathcal{S}_{1}$ to estimate their baseline risk of recurrence, assuming they had not received chemotherapy. This model is referred to as the \textit{X-ray model}, as it provides a prognostic stratification or \say{X-ray} of the data, allowing us to identify discrepancies in the baseline risk between the treatment groups. In an RCT, all treatment groups would have similar baseline risk distributions.

Next, we split the patients into risk buckets based on their recurrence risk estimates $w_{i} = g_{\theta}(\boldsymbol{x}_{i})$ for $i=1, \cdots, n$. Each bucket $k \in \mathcal{K}$ is defined by an interval $[\underbar{w}_{k}, \overline{w}_{k}]$ of baseline recurrence risks, e.g., 10\% to 20\% risk of recurrence, with each patient $i$ assigned to the bucket $k$ where $\overline{w}_{k}$ falls within this range.
The buckets are defined such that each patient belongs to at most one bucket. The thresholds for these buckets, as well as the number of patients in each bucket, depend on the problem and dataset.

To create a cohort that closely resembles the target trial, we propose a modified matching algorithm. This algorithm generates treatment groups with similar baseline risk distributions, emulating random treatment assignment within each bucket. In addition, the risk distribution of the selected patients matches the risk distribution of patients in the trial on average, while the distribution of covariates in the selected cohort should match the distribution of covariates in the trial cohort.

Suppose that $\mathcal{S}_{0}^{sel}$ and $\mathcal{S}_{1}^{sel}$ represent the cohorts of the selected untreated and treated patients after matching, respectively. The matching is designed such that $\sum_{j \in \mathcal{S}_{0}^{sel}} w_{j} = \mu_{0}$ and $\sum_{i \in \mathcal{S}_{1}^{sel}} w_{i} = \mu_{0}$, where $\mu_{0}$ is the trial outcome mean for the group, which is used to construct the model estimating the baseline risk. In this case, $\mu_{0}$ corresponds to the 5-year DFS rate for the untreated group in the RCT. Simply put, the mean baseline risk of recurrence for both the selected untreated and treated patients is set to be the same as the trial mean. In addition, we require that $\sum_{i \in \mathcal{S}_{1}^{sel}} x_{il} = \mu_{1}^{l}$ and $\sum_{j \in \mathcal{S}_{0}^{sel}} x_{jl} = \mu_{0}^{l}$, where $\mu_{a}^{j}$ is the mean value of the $l$th covariate for treatment group $t=a$, as reported in the trial. This ensures that the mean value of each covariate in the selected cohort matches the trial mean for that covariate, assuming that the covariate was reported in the trial.

For example, in our scenario, the trial reported a DFS rate of 38.7\% for the untreated group. Since the treatment assignment in the trial is random, the DFS rate for the treated group would also be 38.7\% had they not received treatment. In addition, the trial reported that 79\% of the patients had left-sided tumors, so $\sum_{i \in \mathcal{S}_{0}^{sel}} x_{ij} = 0.79$, where $x_{ij}$ is a binary covariate that is equal to 1 if the patient has a left-sided tumor and 0 otherwise.

We now describe the optimization algorithm used for the matching. First, we pre-specify the number of patients $n_{k}^{sel}$ that will be matched in each bucket $k \in \mathcal{K}$. This number depends on the data available, such as the ratio of treated to untreated patients in each bucket. The selection is typically configured so that the mean baseline risk distribution of the matched cohort closely matches the trial target for the untreated group. We assume that $\mathcal{S}_{1}$ and $\mathcal{S}_{0}$ contain the treated and untreated patients in the original cohort, respectively. Similarly, $\mathcal{S}_{1}^{k}$ and $\mathcal{S}_{0}^{k}$ contain the treated and untreated patients in the original cohort that belong to bucket $k \in \mathcal{K}$, respectively. Also, $k_{i}$ denotes the bucket of patient $i$. The formulation is

\begin{equation}
    {\everymath{\displaystyle}
    \begin{array}{lll}
         \min_{\boldsymbol{z}} & \left|\mu_{0} - \frac{\sum_{i \in \mathcal{S}_{1}} \sum_{j \in \mathcal{S}_{0}} w_{i} z_{ij}}{\sum_{i \in \mathcal{S}_{1}} \sum_{j \in \mathcal{S}_{0}} z_{ij}} \right|
         + \left|\mu_{0} - \frac{\sum_{i \in \mathcal{S}_{1}} \sum_{j \in \mathcal{S}_{0}} w_{j} z_{ij}}{\sum_{i \in \mathcal{S}_{1}} \sum_{j \in \mathcal{S}_{0}} z_{ij}} \right| \\
         & + \sum_{l=1}^{L} \left|\mu_{1}^{l} - \frac{\sum_{i \in \mathcal{S}_{1}} \sum_{j \in \mathcal{S}_{0}} x_{il} z_{ij}}{\sum_{i \in \mathcal{S}_{1}} \sum_{j \in \mathcal{S}_{0}} z_{ij}} \right|
         + \sum_{l=1}^{L} \left|\mu_{0}^{l} - \frac{\sum_{i \in \mathcal{S}_{1}} \sum_{j \in \mathcal{S}_{0}} x_{jl} z_{ij}}{\sum_{i \in \mathcal{S}_{1}} \sum_{j \in \mathcal{S}_{0}} z_{ij}} \right| \\
         & +  \sum_{i \in \mathcal{S}_{1}}  \sum_{j \in \mathcal{S}_{0}} z_{ij} \| \boldsymbol{x}_{i} - \boldsymbol{x}_{j} \|^{2}  \\
         \text{s.t.} & \sum_{j \in \mathcal{S}_{0}} z_{ij} \leq 1, & \hspace{-95pt} \forall i \in \mathcal{S}_{1}, \\
         & \sum_{i \in \mathcal{S}_{1}} z_{ij} \leq 1, & \hspace{-95pt} \forall j \in \mathcal{S}_{0}, \\
         & \sum_{i \in \mathcal{S}_{1}^{k}} \sum_{j \in \mathcal{S}_{0}} z_{ij} = n_{k}^{sel}, \\
         & z_{ij} = 0, & \hspace{-95pt} \forall i \in \mathcal{S}_{1}, \forall j \in \mathcal{S}_{0}, k_{i} \neq k_{j}, \\
         & z_{ij} \in \{0, 1\}, & \hspace{-95pt} \forall i \in \mathcal{S}_{1}^{k}, \; \; j \in \mathcal{S}_{0}^{k}.
    \end{array}
    }
    \label{eq:eq1}
\end{equation}

The variable $z_{ij}=1$, if patient $i$ is matched to patient $j$; otherwise $z_{ij}=0$. Once we obtain the solution to Problem $\eqref{eq:eq1}$, we only retain the matched patients in our final dataset.

The objective function consists of three key components. The first two terms aim to minimize the absolute differences in the mean baseline risk (e.g., the 5-year recurrence-free survival rates) between the matched cohort and the trial cohort. The subsequent two terms minimize the differences in the means of covariates for the matched cohort, ensuring they align with the covariate distribution in the trial data. Notably, we match the target for each treatment group separately to ensure that both the overall cohort and each individual treatment group are distributed similarly to the trial cohort. The final term minimizes the distance between matched pairs based on selected covariates (e.g., tumor size and tumor site), similar to the matching strategy used in the R.O.A.D. framework.

The constraints enforce the structure of the matching process. In the first constraint, we require that each treated patient is matched with at most one untreated patient. Similarly, the second constraint ensures that each untreated patient is matched with at most one treated patient. The third constraint stipulates that we match $n_{k}^{sel}$ treated patients and $n_{k}^{sel}$ untreated patients from each bucket $k$. It is important to note that the matching algorithm selects the $n_{k}^{sel}$ patients based on the pairs of patients that best fit the goals outlined in the objective function. Finally, the last constraint prohibits matching patients from different buckets.

\subsection{Cost-Sensitive Counterfactual Models and Reward Estimation}

In this step, we utilize direct reward estimation to predict the outcome for each patient under each treatment option. For example, we train a counterfactual model on patients who underwent hepatectomy alone to predict the 5-year RFS under hepatectomy alone. Similarly, we train a model on patients in the hepatectomy plus chemotherapy group. For each patient in the matched cohort, we use each model to estimate the RFS under both treatment options. The mean of these predictions across all patients provides the estimated 5-year RFS for each treatment option, which can then be compared to the DFS reported in the trial. We further adjust for any discrepancies between the estimated RFS and the trial DFS using cost-sensitive learning, extending the approach we pioneered in the R.O.A.D. framework. For example, if the estimated RFS under hepatectomy plus chemotherapy is 44\%, deviating from the trial target of 49.5\%, we assign larger weights to the patients in the hepatectomy plus chemotherapy group who did not experience a recurrence, thereby increasing the estimated treatment effect of hepatectomy plus chemotherapy. Of note, the weight tuning process is straightforward in this case, as we gradually adjust the weights until the estimated RFS aligns with the trial target.

Specifically, we train two new classification models to estimate the 5-year RFS probability: one model, $h_{t=1}(\boldsymbol{x})$, in the matched treated group, and another, $h_{t=0}(\boldsymbol{x})$, in the matched untreated group. This approach, known as \say{direct} reward estimation, is used to evaluate different policies or treatments \cite{dudik2011doubly}. It is important to note that training both models in the matched cohorts, with an identical number of patients in each bucket, ensures they \say{focus} on the same patient subgroups. We then compare the mean RFS estimates $\hat{h}_{t=0} = \frac{1}{n^{sel}} \sum_{i \in (\mathcal{S}_{1}^{sel} \cup \mathcal{S}_{0}^{sel})} h_{t=0}(\boldsymbol{x}_{i})$ and $\hat{h}_{t=1} = \frac{1}{n^{sel}} \sum_{i \in (\mathcal{S}_{1}^{sel} \cup \mathcal{S}_{0}^{sel})} h_{t=1}(\boldsymbol{x}_{i})$ for the two treatment options across all $n^{sel}$ selected patients. 

Due to the objective of the matching algorithm, we expect $\hat{h}_{t=0}$ to closely approximate the trial target $\mu_{0}$, e.g., the estimated 5-year RFS rate under no treatment should align with the trial's control group RFS rate. Morevoer, we ensure that the selected treated patients have a similar baseline prognosis, so we expect $\hat{h}_{t=1}$ to be close to the trial target $\mu_{1}$, e.g. the estimated 5-year RFS rate for the treated group. However, due to limitations in the data, such as unobserved confounding or inaccuracies in the prognostic models, $\hat{h}_{t=1}$ may underestimate the effect of treatment. To correct for this and align $\hat{h}_{t=1}$ with the target trial $\mu_{1}$, we re-train the counterfactual model using a cost-sensitive loss function. This function assigns larger weights to patients who received treatment and had positive outcomes, thereby improving predictions for that group and shifting the RFS rate toward the RFS rate of treated group in the target trial.

More formally, we train the classification models $h_{t=0}(\boldsymbol{x})$ and $h_{t=1}(\boldsymbol{x})$ by minimizing a loss function $\ell$, such as the cross-entropy loss. Let $\mathcal{S}_{1}^{-}$, $\mathcal{S}_{1}^{+}$ represent the set of patients in $\mathcal{S}_{1}^{sel}$ who did or did not experience a recurrence, respectively. To train the classification model $h_{t=1}(\boldsymbol{x})$, we use a modified empirical loss function:
\begin{equation}
 \label{1trainng}
    {\everymath{\displaystyle}
        \sum_{i \in \mathcal{S}_{1}^{-}} \ell(y_{i}, h_{t=1}(\boldsymbol{x}_{i})) + \rho \sum_{j \in \mathcal{S}_{1}^{+}} \ell(y_{j}, h_{t=1}(\boldsymbol{x}_{j})).
    }
\end{equation}
Suppose we increase the weight $\rho$ for the patients who received chemo but did not have a recurrence. The model will then try to make better predictions for the patients with the larger weight, since this yields a greater improvement in the loss \cite{elkan2001foundations}.

Using the weighted loss, the RFS estimates of $h_{t=1}(\boldsymbol{x})$ also depend on the weight $\rho$. To make this dependence explicit, we use the notation $h_{t=1}(\boldsymbol{x},\rho)$. In this case, we expect as $\rho \geq 1$ increases, $\hat{h}_{t=1}$ to increase. By adjusting the weight such that $\hat{h}_{t=1} = \mu_{1}$, we eliminate the effect of the unobserved confounding.

\subsection{Optimal Policy Trees}
\label{sec:opts}

The first three steps of the proposed approach will result in an emulated cohort of patients that closely mirrors the target RCT. The control and treatment groups in the emulated cohort will have a similar distribution of patient characteristics and baseline risk of recurrence as the trial, while the counterfactual RFS estimates of each treatment group will match those reported in the target trial. With this, we can apply a methodology to identify subgroups of patients with distinct characteristics who exhibit heterogeneity in treatment effects (HTE).

To achieve this, we will utilize the probabilities derived from the two counterfactual models to train an Optimal Policy Tree (OPT) with direct reward estimation \cite{amram2022optimal}. This allows us to group patients with similar HTE within each node of the tree. 

The tuning of various hyperparameters allows the training of multiple OPTs. The criterion for selecting the final OPT is partly based on maximizing concordance between the optimal treatment predicted by the counterfactual models for each patient and the treatment assigned by the OPT. In other words, if we have a cohort of 100 patients and the counterfactual models predict that 60 of them would benefit from a particular treatment, we would select the OPT that assigns the highest number of these 60 patients to the nodes that recommend that treatment. Before using concordance as the criterion for OPT selection, however, we must ensure that the tuning hyperparameters are appropriate for the dataset at hand. For example, the sample size of the training cohort is crucial, as it will determine the minimum number of patients in each node. A very small number of patients per node may lead to higher concordance but at the cost of overfitting.

\subsection{Constrained Reward Estimation}

Since the OPTs aim to assign the treatment option that would yield the highest RFS for each individual patient, they can allocate specific treatments even when the treatment effect is small. 

In cases where we emulate an overall negative trial and wish to identify subsets of patients who benefit from the treatment, we can use constraints to ensure that the treatment effect exceeds a pre-specified threshold, for each subgroup recommended the treatment by the OPT. This threshold represents the clinically significant treatment effect for the specific clinical problem.

Conversely, when emulating an overall positive trial, the goal is to identify subgroups that do not benefit from treatment. In line with the Hippocratic principle of \say{first, do no harm} we can apply constraints that minimize the likelihood of the OPT denying an effective treatment to any subset of patients.

These constraints can be incorporated into the OPT formulation. In the case of a negative target trial, the constraints would decrease the mean rewards, e.g. RFS rate, under treatment to prevent the OPT from making further splits if the mean treatment effect in any resulting group falls below the specified threshold. Conversely, for a positive target trial, such as the JCOG0603 trial used in our demonstration, the constraints would decrease the mean rewards under no treatment to increase the likelihood that OPT subsets are assigned treatment. Since these constraints are not directly part of the OPT implementation, we approximate them by adjusting the rewards.

In this demonstration, for patients with higher rewards under no chemotherapy compared to chemotherapy, we decrease their rewards by setting them to 0.78 of the value under chemotherapy. This was the largest constraint we could apply while still allowing the OPT to split. In other words, if we set a greater constraint, all patients would have been assigned adjuvant chemotherapy by the OPT.

\begin{algorithm}[]
	\caption{The R.O.A.D. to randomized clinical trial emulation. An example of observational data of patients with CRLM with treatment and no treatment options.} 
	\label{alg:alg_1}
	\SetKwComment{Comment}{/* }{*/}
	\SetAlgoLined 
	\KwIn{ \SetKwData{}{} \\
            \text{Original dataset} ($\boldsymbol{X}$, $\boldsymbol{t}$, $\boldsymbol{y}$) \\
            \text{Target RCT, including outcome and covariate means} \\
	\textbf{Parameters:} \SetKwData{}{} \\
	\begin{itemize}
	    \item $\mathcal{K}$: set of risk buckets
	    \item $[\underbar{$w$}_{k}, \overline{w}_{k}]$: risk bucket thresholds for $k=1, \cdots, m$
	    \item $g_{\theta}$: ML model for predicting the baseline risk of each patient (default: RF)
            \item $h_{t=0}, \; \; h_{t=1}$: counterfactual models for predicting the outcome (RFS) under each treatment (default: RF) 
            \item $\rho$: weight of the patient group of interest
	    \item $OPT$: Optimal Policy Tree that optimally assigns        treatments to subgroups of patients
	\end{itemize}} 
	\KwOut{Trained $OPT$ and matched cohort}
    \Comment{Step 1: Inclusion trial criteria}
    Filter out ($\boldsymbol{X}$, $\boldsymbol{t}$, $\boldsymbol{y}$) based on the inclusion criteria \\
    \Comment{Step 2: Risk Stratification and Matching}
    Train $g_{\theta}$ on the untreated patients $(\boldsymbol{X}, \boldsymbol{t}=0, \boldsymbol{y})$ in $(\boldsymbol{X}, \boldsymbol{t}, \boldsymbol{y})$ \\
    \Comment{Select the patients using the proposed matching algorithm}
    $(\boldsymbol{X}, \boldsymbol{t}, \boldsymbol{y}) = sel(\boldsymbol{X}, \boldsymbol{t}, \boldsymbol{y})$ \\
    \Comment{Step 3: Cost-sensitive Counterfactual Estimation}
    Train $h_{t=0}$ on $(\boldsymbol{X}, \boldsymbol{t}=0, \boldsymbol{y})$ \\
    Train $h_{t=1}(\cdot, \rho)$ on $(\boldsymbol{X}, \boldsymbol{t}=1, \boldsymbol{y})$ with $\rho=1$ \\
    $\hat{h}_{t=0} = \frac{1}{n^{sel}} \sum_{i \in (\mathcal{S}_{1}^{sel} \cup \mathcal{S}_{0}^{sel})} h_{t=0}(\boldsymbol{x}_{i})$, the mean rewards or estimated RFS under $t=0$ \\
    $\hat{h}_{t=1} = \frac{1}{n^{sel}} \sum_{i \in (\mathcal{S}_{1}^{sel} \cup \mathcal{S}_{0}^{sel})} h_{t=1}(\boldsymbol{x}_{i})$, the mean rewards estimated RFS under $t=1$ \\
    \Comment{Gradually increase or decrease the weight until the mean rewards match the trial target}
    Obtain optimal $\rho$ \\
    \Comment{Step 5: Train OPT}
    Train $OPT$ on the counterfactual estimations $(\boldsymbol{x}_{i}, h_{t=0}(\boldsymbol{x}_{i}))$ and $(\boldsymbol{x}_{i}, h_{t=1}(\boldsymbol{x}_{i}, \rho))$ for all $\boldsymbol{x}_{i}$ in $\boldsymbol{X}$ \\
    \Comment{Step 6: Constrained Rewards}
    Adjust rewards to obtain an OPT where the HTE of each subgroup is above the pre-specified value \\
    \Return $OPT$ 
\end{algorithm} 

\subsection{Validation of Absence of Confounding}
\label{sec:val}

We next present a method for assessing the presence of observed confounding at each subgroup or node of the OPT. Since we did not employ techniques such as matching to reduce observed confounding within the tree nodes, the absence of observed confounding in these nodes would provide reasonable evidence that unobserved confounding is also absent-similar to how the balance of measured baseline characteristics in an RCT indicates successful randomization. While this step is not strictly necessary - the validation through Kaplan Meier (KM) plots in both the internal observational and external randomized controlled trial cohorts will ultimately validate the framework - it is still informative to evaluate potential confounding within the tree nodes. To evaluate confounding, we utilize prognostic models, similar to the baseline model $g_{\theta}(\boldsymbol{x})$ in Step 2 of the framework, with the main difference being that these new models are trained on external cohorts to ensure no information leakage. These models are then applied to each subgroup of patients to obtain the estimated outcome for each individual. Similar to the baseline model, the prognostic models are counterfactuals, since they provide predictions under the same treatment for all patients. If the prognostic models provide similar estimated outcomes for the patients in each treatment group at a specific node, there is minimal confounding in that node and the differences in the outcomes can be attributed mostly to the treatment.

In the CRLM example, we calculated the 5-year RFS for patients receiving surgery plus chemotherapy versus those undergoing surgery alone, within each node. This was done using two external prognostic models: the Memorial Sloan Kettering (MSK) Clinical Risk Score (CRS) for colorectal liver metastases (CRLM), and the Johns Hopkins Hospital (JHH) Genetic And Morphological Evaluation (GAME) score for CRLM \cite{fong1999clinical, margonis2018genetic}. The MSK CRS score is calculated based on the following factors: one point for primary tumor lymph node metastasis, one point for a disease-free interval from primary to metastases of less than 12 months, one point for having more than one hepatic tumor, one point for a largest hepatic tumor greater than 5 cm, and one point for carcinoembryonic antigen (CEA) $\ge$ 200 ng/mL. The total MSK CRS score ranges from 0 to 5. The JHH GAME score is calculated as follows: one point for KRAS-mutated tumors, one point for CEA $\geq$ 20 ng/mL, one point for primary tumor lymph node metastasis, one point for a Tumor Burden Score (TBS) between 3 and 9, and two points for a TBS $\geq$ 9. The total GAME score ranges from 0 to 5. These models were trained on external datasets, ensuring no information leakage.

Ideally, we expect the MSK CRS and JHH GAME model predictions of baseline recurrence risk for patients treated with surgery alone to closely align with the predictions for patients that received surgery plus chemotherapy. This consistency would further confirm that confounding has been minimized.

\subsection{Comparison of Actual Treatment Effect: Emulated Cohort vs. Target RCT Cohort}
\label{sec:comp}

To evaluate the actual treatment effect, we utilize the same endpoints as the trial and report the outcomes for each subgroup or node of the OPT. In the CRLM demonstration, we validated the framework by comparing the actual effect of adjuvant chemotherapy - determined by Kaplan-Meier plots - across subsets defined by the OPT in both the emulated cohort and the RCT cohort. Specifically, we compared the direction of the treatment effect (i.e., benefit vs. no benefit), the size of the benefit (i.e., the difference in median RFS between untreated and treated patients among those who benefit), and the statistical significance of this benefit. For successful validation, the direction of the treatment effect must be the same in the emulated and target RCT cohorts for each subset defined by the OPT. If there is a statistically significant benefit in a specific subset of the emulated cohort, the same should be true for the target RCT cohort. Ideally, the size of this effect should be very similar between the two cohorts. Lastly, if the validation is successful, we expect the proportion of patients benefiting from the treatment to be similar between the emulated cohort and the target RCT cohort. Note that in the application of the proposed framework, we do not need to validate the treatment recommendations in the target RCT cohort, especially since the target RCT cohort is not typically available. However, in this demonstration of our framework, we provide the target RCT validation to offer stronger evidence for our approach.

Note that in the application of the proposed framework, we do not need to validate the treatment recommendations in the target RCT cohort, especially since the target RCT cohort is not typically available. Internal validation of the OPT recommendations, by comparing the actual treatment effect-determined via Kaplan-Meier plots-across subsets defined by the OPT in the emulated cohort, is sufficient. However, in this demonstration of our framework, we include validation using the target RCT cohort to provide stronger evidence for our approach.

\section{Results}

\subsection{Real-world Dataset vs. JCOG0603 RCT}

The real-world dataset consisted of observational data from 779 adult patients who underwent curative-intent resection of CRLM at the tertiary centers participating in the IGCLM consortium between 2002 and 2019, and who met the eligibility criteria of the JCOG0603 trial. Of these, 639 patients underwent surgery alone, while 140 patients underwent surgery in combination with adjuvant chemotherapy.

The baseline risk distribution of this real-world cohort is illustrated in Figure 1, showing a considerable imbalance in the baseline recurrence risk between patients who received adjuvant chemotherapy and those who did not. Additionally, we compared the distribution of the four prognostically important covariates reported in the original RCT - namely, the number of patients with metastatic lymph nodes, the number of patients with $\geq$4 liver metastases, the number of patients with a liver metastasis $\geq$5 cm, and the number of patients with right versus left-sided liver metastases-between the real-world cohort and the JCOG0603 trial. The most notable difference was in the frequency of patients with metastatic lymph nodes, with 60\% in the real-world dataset compared to 40\% in the RCT cohort.

\begin{figure}[!h]
\centering
\caption{Distribution of baseline recurrence risk in (A) the pre-prognostic matching cohort and (B) the post-prognostic matching cohort}
\includegraphics[width=1\linewidth]{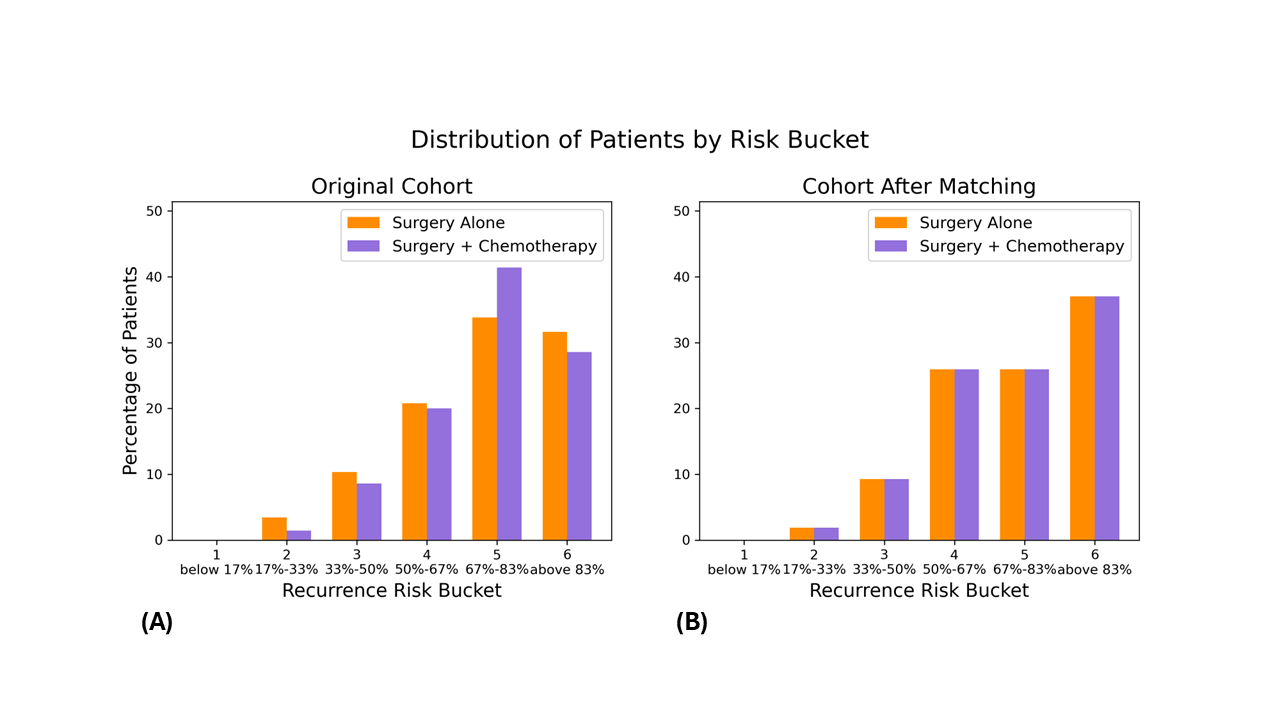}
\end{figure}

\subsection{Prognostic matching}

The 5-year RFS Kaplan-Meier estimate for the untreated group in the trial was 39\%, with the only prognostically important variables reported in the original RCT being the number of patients without versus with metastatic lymph nodes, the number of patients with 1-3 versus $\geq$ 4 liver metastases, the number of patients with a liver metastasis $<$ 5 cm versus $\geq$ 5 cm, and the number of patients with right versus left-sided liver metastases. We used this information to solve the optimization problem outlined in the methods section 2.2. This resulted in deriving an emulated cohort of 216 patients (108 treated with surgery alone and 108 treated with surgery plus adjuvant chemotherapy) with a predicted baseline 5-year RFS of 35\%, closely matching that of the trial cohort. It also resulted in a cohort with balanced covariates between the two treatment groups, as well as a distribution similar to that of the target trial (Table 1).

\begin{table}[!h]
\caption{Demographics and Baseline Characteristics of the Emulated Cohort.}
\normalsize 
\begin{tabularx}{\textwidth}{|X||X|}
\hline
\textbf{Characteristic}  & \textbf{ (n=216)} \\ \hline
\textbf{Age (median [IQR])} & 62.55[53.64, 69.00]\\ \hline
\textbf{Sex (n, \%)} & \\ \hline
     Male	&156(72) \\ \hline
      Female&	60(28)\\ \hline
   \textbf{T stage (n, \%)} & \\ \hline
      0&	1(0)\\ \hline
   1&	8(4)\\ \hline
   2	&36(17)\\ \hline
   3	&134(62)\\ \hline
   4	&37(17)\\ \hline
   \textbf{N status (\%)} & \\ \hline
      Positive&	100(46) \\ \hline
   Negative	&116(54) \\ \hline
   \textbf{Bilobar distribution (\%)}  &\\ \hline
      Yes	&41(19)\\ \hline
   No	&175(81)\\ \hline
   \textbf{Size of largest liver metastasis equal or greater than 5 cm}  &\\ \hline
      Yes	&31(14)\\ \hline
   No	&185(86)\\ \hline
   \textbf{Number of liver metastases equal or greater than 4}  &\\ \hline
    Yes&	14(6)\\ \hline
   No	&202(94)\\ \hline
      \textbf{CEA more than 20 ng/mL} & \\ \hline
    Yes	&61(28)\\ \hline
   No	&155(72)\\ \hline
      \textbf{Metachronous presentation} & \\ \hline
     Yes&	82(38)\\ \hline
   No&	134(62)\\ \hline
        \textbf{Left-sided tumor} &\\ \hline
   Yes	&170(79)\\ \hline
   No&	46(21)\\ \hline
\end{tabularx}
\end{table}

\subsection{Weight tuning}

In the emulated cohort described above, the median follow-up was 72.03 (IQR: 50.66;94.56) months. Patients treated with surgery alone had a better 5-year KM RFS (38\% vs. 34\%; p = 0.016). Importantly, the 5-year KM RFS for the group that received adjuvant chemotherapy was 11\% lower than that observed in the target trial (49\%). This underestimation of the benefit of adjuvant chemotherapy reflects the effect of unobserved confounding. The challenge with unobserved confounding is that we cannot match for it, as these factors are not observed or measured. However, using this framework, we applied a weight of 1.5 for treated patients with no recurrence, which improved the predicted 5-year RFS, now matching that reported in the treated group of the target RCT.

The weight-tuning tool is also useful in situations where prognostic matching is not entirely successful. For instance, in cases where important prognostic factors, such as disease-free interval (DFI) or tumor location (as demonstrated in our study), are not included in the RCT's published Table 1, making it impossible to match these characteristics. Additionally, weight tuning can help when we cannot fully match certain patient characteristics or emulate the control group's prognosis due to sample size limitations. By treating these unreported factors as unobserved confounders, we can use weight tuning to adjust for their impact and improve the validity of our emulation. In this demonstration, we applied a weight of 1.25 for untreated patients with no recurrence, which improved the predicted 5-year RFS, now aligning it with that reported in the untreated group of the target RCT.

\subsection{OPT 1}

The emulated cohort was used to train random survival forest (RF) counterfactual models using the following variables: patient age, gender, T stage, N stage, tumor laterality (right vs. left), CEA, disease-free interval (DFI), bilobar involvement, tumor size, and number of metastases. The counterfactual model achieved an average C-index of 0.88 in-sample, with c-indices of 0.68 for untreated patients and 0.61 for treated patients, as evaluated by 3-fold cross-validation. The first OPT (Figure 2A) categorized the trial cohort into five subsets based on distinct benefits from the addition of adjuvant chemotherapy. Among these, three subsets (nodes: 2, 8, and 9), comprising 186 out of 216 patients, were identified as benefiting from chemotherapy. The first subset consisted of patients with a DFI less than 12.5 months. The second subset included those with a DFI equal to or greater than 12 months and a tumor size equal to or greater than 4.6 cm. The third subset included patients with a DFI equal to or greater than 12.5 months, a T stage of 3 or 4, and a tumor size between 2.25 and 4.6 cm. The fact that two of these nodes did not show an actual benefit upon internal and external validation is not problematic, as these nodes would simply not be used based on the interval validation results. The fact that two of these nodes were not shown to have an actual benefit on internal validation (and external validation) is not problematic as we would just not use these nodes in the first place based on the interval validation. The two subsets (nodes: 5 and 7) that did not benefit from adjuvant chemotherapy were: 1) patients with a DFI equal to or greater than 12 months, a tumor size less than 4.6 cm, and either a T stage of 1 or 2, and 2) patients with a DFI equal to or greater than 12 months, a tumor size less than 2.25 cm, and a T stage of 3 or 4. 

\begin{figure}[!h]
\centering
\caption{(A) First OPT, (B) Second OPT.}
\includegraphics[width=1\linewidth]{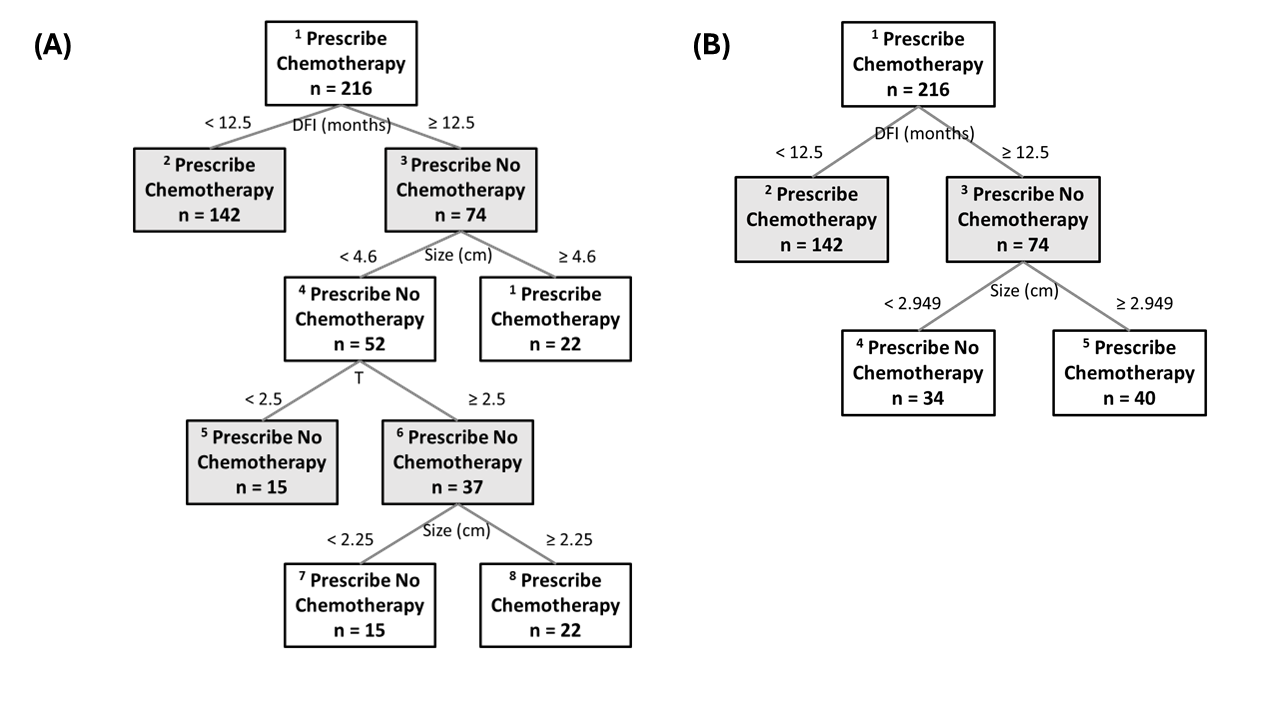}
\end{figure}

\subsection{OPT 2}

The second OPT (Figure 2B) classified the trial cohort into three subsets based on distinct benefits from adjuvant chemotherapy. Among these, two subsets (nodes: 2 and 5), comprising 182 out of 216 patients, were identified as benefiting from chemotherapy. The first subset included patients with a DFI less than 12.5 months, and the second subset included those with a DFI equal to or greater than 12 months and a tumor size equal to or greater than 2.95 cm. The subset that did not benefit from adjuvant chemotherapy consisted of patients with a DFI equal to or greater than 12 months and a tumor size less than 2.95 cm.

\subsection{Validation of Absence of Confounding}

To ensure that there was no confounding within the five subsets defined by the first OPT, we analyzed the baseline risk of recurrence, as reflected in the predictions from the MSK CRS and the JHH GAME score. As shown in Table 2, the risk-according to either external prognostic model-was equally distributed between patients who received adjuvant chemotherapy and those who did not, within each of the five subsets. Therefore, we concluded that there was no confounding, and the observed differences in 5-year recurrence-free survival within each subset can be attributed to the effect of adjuvant chemotherapy, rather than any confounding factors, whether observed or unobserved.

\begin{table}[!h]
\begin{threeparttable}
\caption{The mean MSK CRS and JHH GAME scores for the nodes of the first OPT.}
\normalsize 
\begin{tabularx}{\textwidth}{|X||X|X|X|}
\hline
\textbf{Node}  & \textbf{Surgery alone} & \textbf{Adjuvant chemotherapy} & \textbf{P-value}\\ \hline
\textbf{Node 2} & & & \\ \hline
   MSK CRS & 1.929 & 2.112 &	0.269 \\ \hline
   JHH GAME & 1.591 & 1.901 &	0.070\\ \hline
\textbf{Node 5} & & &\\ \hline
   MSK CRS & 0.625	& 0.571 &	0.846 \\ \hline
   JHH GAME & 1.250&	1.428	&0.688\\ \hline
   \textbf{Node 7}& & & \\ \hline
   MSK CRS & 0.714	&0.500	&0.434 \\ \hline
   JHH GAME & 0.857&	1.125	&0.595\\ \hline
   \textbf{Node 8}& & & \\ \hline
   MSK CRS & 0.818&	0.363&	0.111 \\ \hline
   JHH GAME & 1.545	&1.727	&0.570\\ \hline
   \textbf{Node 9} & & &\\ \hline
   MSK CRS & 1.000&	1.181	&0.431 \\ \hline
   JHH GAME & 1.909	&2.272&	0.393\\ \hline
\end{tabularx}
\begin{tablenotes}
      \footnotesize
      \item[*]The Memorial Sloan Kettering (MSK) Clinical Risk Score (CRS) for colorectal liver metastases (CRLM), and the Johns Hopkins Hospital (JHH) Genetic And Morphological Evaluation (GAME) score for CRLM.
    \end{tablenotes}
\end{threeparttable}
\end{table}

Similarly, to ensure there was no confounding within the three subsets defined by the second OPT, we analyzed the baseline risk of recurrence as reflected in the MSK CRS and JHH GAME score predictions. As shown in Table 3, the risk was equally distributed between patients who received adjuvant chemotherapy and those who did not, within each of the three subsets. Based on this, we concluded that there was no confounding, and the differences in 5-year RFS within each subset can be attributed to the effect of adjuvant chemotherapy, rather than any confounding factors.

\begin{table}[!h]
\begin{threeparttable}
\caption{The mean MSK CRS and JHH GAME scores for the nodes of the second OPT.}
\normalsize 
\begin{tabularx}{\textwidth}{|X||X|X|X|}
\hline
\textbf{Node}  & \textbf{Surgery alone} & \textbf{Adjuvant chemotherapy} & \textbf{P-value}\\ \hline
\textbf{Node 2} & & & \\ \hline
   MSK CRS & 1.929&	2.112	&0.269\\ \hline
   JHH GAME & 1.591	&1.901	&0.070\\ \hline
\textbf{Node 4} & & &\\ \hline
   MSK CRS & 0.833&	0.437	&0.052 \\ \hline
   JHH GAME & 1.111	&1.312	&0.529\\ \hline
   \textbf{Node 5}& & & \\ \hline
   MSK CRS & 0.789	&0.857&	0.724 \\ \hline
   JHH GAME & 1.789&	2.000	&0.436\\ \hline
\end{tabularx}
\begin{tablenotes}
      \footnotesize
      \item[*]The Memorial Sloan Kettering (MSK) Clinical Risk Score (CRS) for colorectal liver metastases (CRLM), and the Johns Hopkins Hospital (JHH) Genetic And Morphological Evaluation (GAME) score for CRLM.
    \end{tablenotes}
\end{threeparttable}
\end{table}

\subsection{Comparison of Actual Treatment Effect: Emulated Cohort vs. Target RCT Cohort}


The effect of chemotherapy was assessed using the KM product-limit method, which calculates the outcomes of patients who received adjuvant chemotherapy versus those who did not. The KM method reflects the ground truth, as it uses actual outcomes - namely, the presence of an event and the time to event or last follow-up - unlike prognostic models that rely on predictions.

\noindent Validation was performed on three levels:

\begin{enumerate}
  \item Direction (benefit vs. no benefit of adjuvant chemotherapy),
  \item Effect size if chemotherapy was beneficial, and
  \item Statistical significance of chemotherapy benefit, assessed by the log-rank test.
\end{enumerate}

The first comparison examined the effect of chemotherapy in all patients recommended to receive adjuvant chemotherapy by the first OPT, in both the emulated and external validation cohorts. In the emulated cohort, adjuvant chemotherapy benefited these patients by improving their median RFS by 22.2 months, which was statistically significant (median RFS in those who did not receive adjuvant chemotherapy: 10.9 months vs. 33.1 months; p = 0.001) (Figure 3A). In the external validation cohort, a similar benefit in median RFS of 24 months was noted, also statistically significant (median RFS in those who did not receive adjuvant chemotherapy: 16.8 months vs. 40.8 months; p = 0.05) (Figure 3B).

\begin{figure}[!h]
\centering
\caption{Recurrence-free survival for patients who were recommended adjuvant chemotherapy by the first OPT, in (A) the emulated cohort and (B) the RCT cohort.}
\includegraphics[width=1\linewidth]{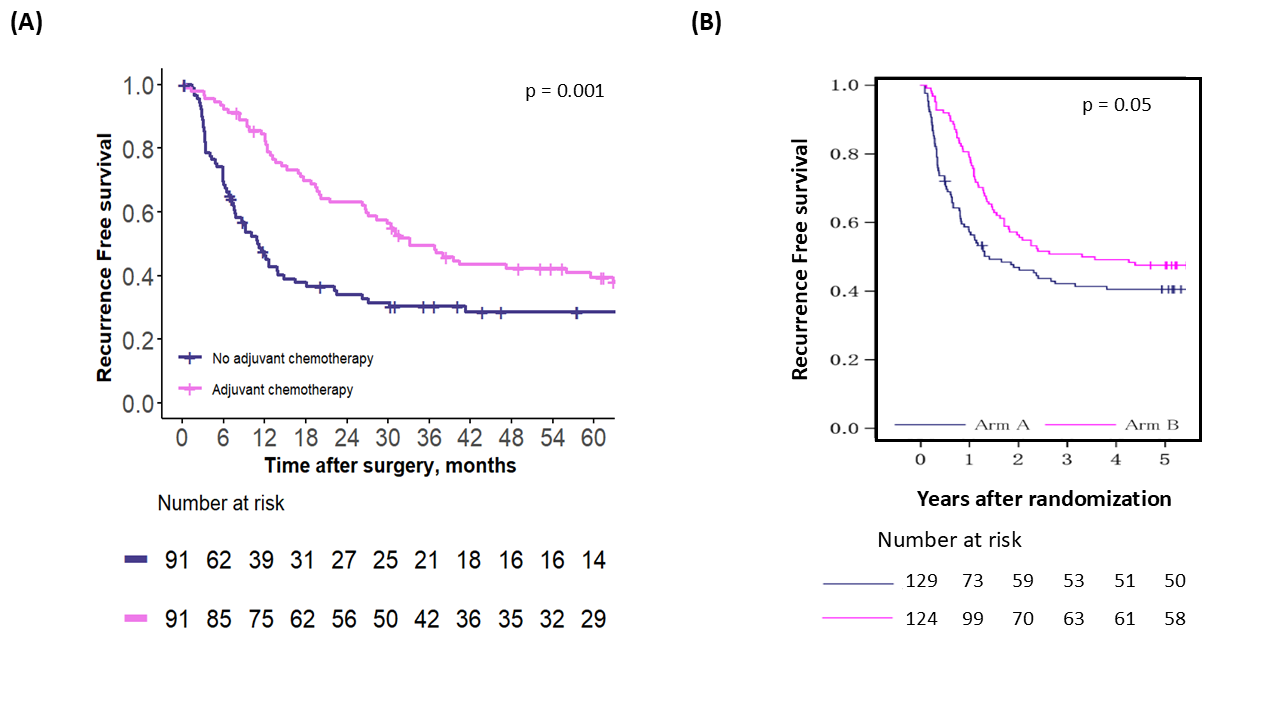}
\end{figure}

The second comparison assessed the effect of chemotherapy in all patients advised against adjuvant chemotherapy by the first OPT, in both the emulated and external validation cohorts. In both cohorts, adjuvant chemotherapy did not benefit these patients (Figure 4A-B, p-values = 0.24 and 0.43, respectively). The comparison between the effect of chemotherapy in the emulated vs. external validation cohorts, for the patient subsets defined by each OPT node, revealed similar effects-or lack of effect-of chemotherapy across the cohorts (eFigures 1-5).

\begin{figure}[!h]
\centering
\caption{Recurrence-free survival for patients who were advised against adjuvant chemotherapy by the first OPT, in (A) the emulated cohort and (B) the RCT cohort.}
\includegraphics[width=1\linewidth]{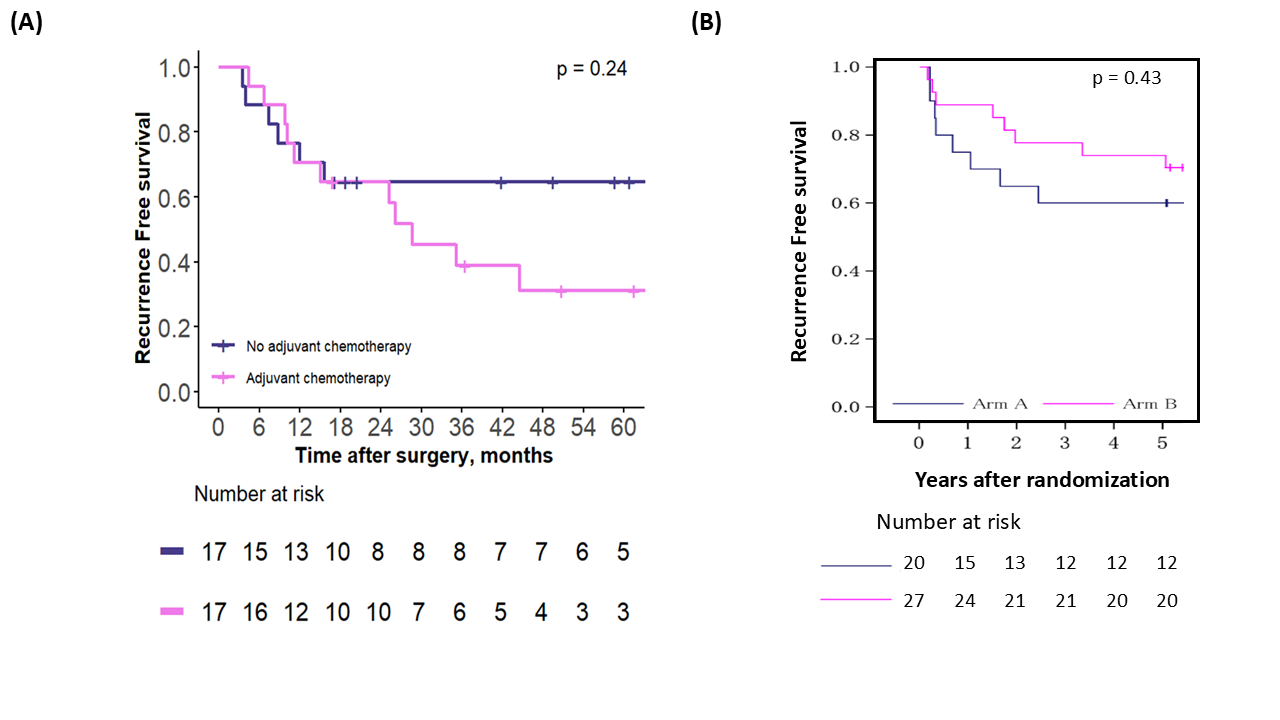}
\end{figure}

The third comparison examined the effect of chemotherapy in all patients recommended to receive adjuvant chemotherapy by the second OPT in both the emulated and external validation cohorts. In the emulated cohort, adjuvant chemotherapy benefited these patients by improving their median RFS by 21.7 months, which was statistically significant (median RFS in those who did not receive adjuvant chemotherapy: 10.1 months vs. 31.8 months; p $<$ 0.001) (Figure 5A). In the external validation cohort, a similar benefit in median RFS of 20.4 months was noted, also statistically significant (median OS in those who did not receive adjuvant chemotherapy: 15.6 months vs. 36 months; p = 0.01) (Figure 5B).

\begin{figure}[!h]
\centering
\caption{Recurrence-free survival for patients who were recommended adjuvant chemotherapy by the second OPT, in (A) the emulated cohort and (B) the RCT cohort.}
\includegraphics[width=1\linewidth]{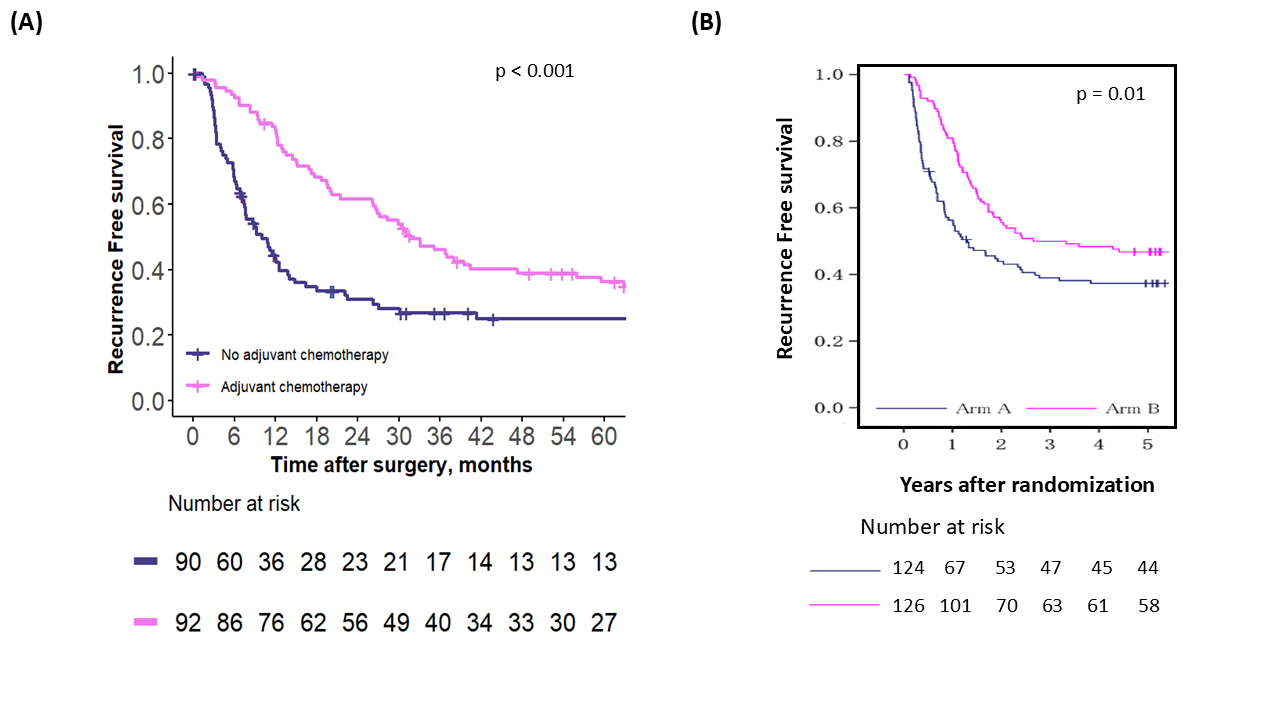}
\end{figure}

The last comparison assessed the effect of chemotherapy in all patients advised against adjuvant chemotherapy by the second OPT, in both the emulated and external validation cohorts. In both cohorts, adjuvant chemotherapy did not benefit these patients (Figure 6A-B, p-values = 0.11 and 0.94, respectively). Similar to the previous comparison, the effects-or lack of effects-of chemotherapy were consistent across both emulated and external validation cohorts (eFigures 6-7).

\begin{figure}[!h]
\centering
\caption{Recurrence-free survival for patients who were advised adjuvant chemotherapy by the second OPT, in (A) the emulated cohort and (B) the RCT cohort.}
\includegraphics[width=1\linewidth]{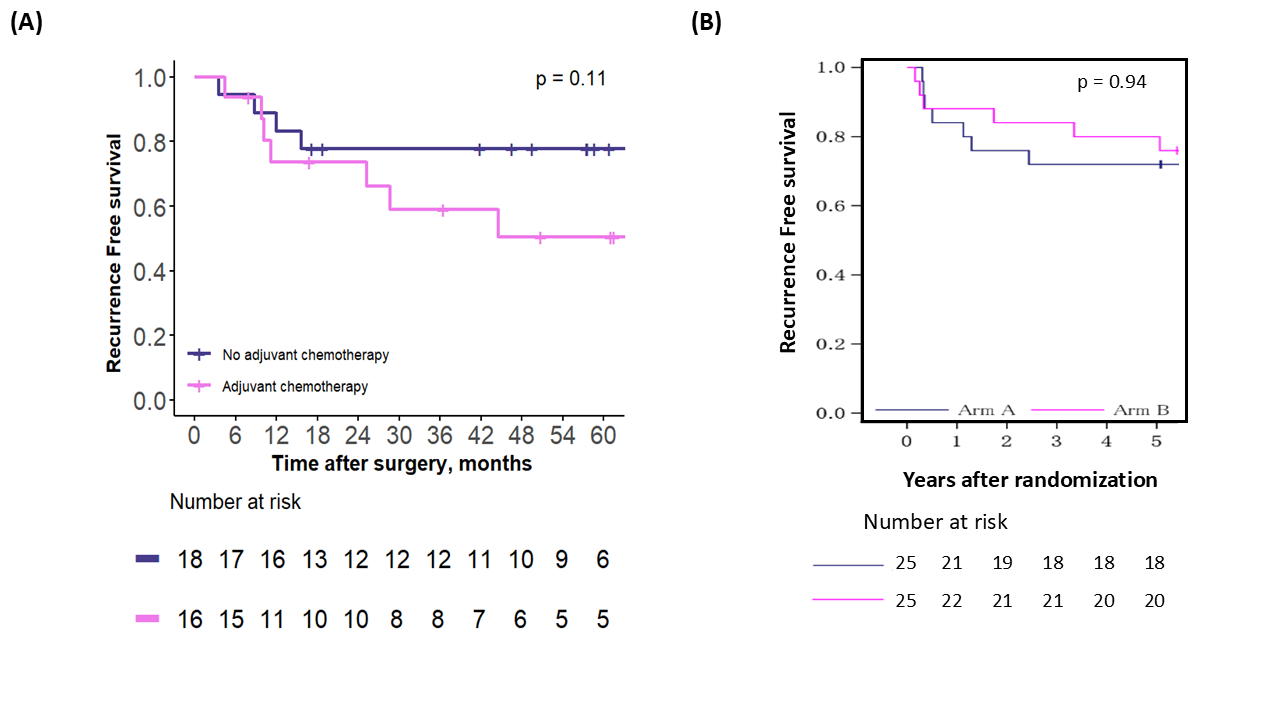}
\end{figure}

Notably, of the 186 patients recommended chemotherapy by the first OPT, 177 were also assigned chemotherapy by the second OPT, resulting in a concordance rate of 95.2\%.

\newpage
 \section{Discussion}

In this work, we present a novel framework for emulating clinical trials using observational data. Emulating clinical trials to guide clinical decision-making is a key objective, and while various approaches have been proposed, our approach introduces four key methodological innovations: (1) a novel prognostic matching technique for extracting a cohort from observational data, which aligns both patient characteristics and baseline prognosis in the treated and untreated groups with those of a target trial; (2) a new weight-tuning method that addresses unobserved confounding by aligning post-treatment prognosis in the treated group with that observed in the target trial; (3) the ability to identify patient subgroups with varying treatment benefits, including those who may experience no benefit at all; and (4) external validation of the framework using an RCT cohort that we did not have prior access to, ensuring unbiased validation. We demonstrate this methodology in the challenging clinical context of colorectal liver metastasis adjuvant treatment, a scenario characterized by limited understanding  of the factors influencing patient outcomes and unclear clinical outcomes despite multiple RCTs \cite{wong2022performance, kokkinakis2024clinical}. This complexity allowed us to rigorously assess whether our methodology can be applied universally, even in technically demanding clinical scenarios.

The first innovation of our approach contrasts with traditional trial emulation methods, which typically assume that applying trial eligibility criteria to an observational cohort ensures a similar distribution of patient characteristics. However, observational datasets often reflect patient selection biases from institutional practices (e.g., differences between academic vs. rural centers) that may differ from those in randomized trials. In contrast, a randomized trial - at least in theory - represents a more unbiased patient selection process. That said, some randomized trials, particularly those with both a placebo arm and a treatment arm involving an established therapy, may have their own selection biases. For example, clinicians may be reluctant to enroll high-risk patients in trials, particularly those with placebo arms, fearing they might be assigned to the placebo arm and thus forgo treatment. This bias would not be captured by the trial's eligibility criteria, meaning that simply applying these criteria to an observational cohort will not guarantee a similar distribution of patient characteristics in the emulated cohort. Our findings support this argument. When we applied the eligibility criteria from the JCOG0603 trial to our observational cohort, we observed significant differences in the proportion of patients with metastatic lymph nodes - a key prognostic factor. Specifically, 61\% of untreated patients in our cohort had metastatic lymph nodes, compared to 40\% in the trial cohort. This discrepancy contributed to a lower 5-year RFS KM estimate for the untreated group in our cohort (27\%), compared to 39\% in the trial cohort. However, by using our prognostic matching technique, we aligned the distribution of patient characteristics and the baseline prognosis in both groups, improving the 5-year RFS KM estimate for the untreated group in our cohort (34\%), and bringing it much closer to the trial's results for the untreated group.

Emulating the 5-year RFS KM estimate for the untreated group of the target trial posed particular challenges, as this required the use of a counterfactual model, and predicting prognosis in CRLM is notoriously difficult \cite{margonis2022precision}. Despite the existence of numerous prognostic models, none has demonstrated excellent discriminatory capacity, reflecting the limited understanding of the factors influencing patient outcomes. Similarly, our survival model, had a cross-validated C-index of only 0.65. How, then, did we successfully emulate the trial's baseline DFS? The answer lies in our model's in-sample C-index, which was excellent at 0.88, indicating the model's ability to \say{memorize} the data within the study sample. While predictive performance across a broader population may vary, the high in-sample concordance was sufficient for our framework's purposes. This was reflected in our model's predicted 35\% 5-year RFS, which was nearly identical to the 34\% estimate derived from Kaplan-Meier analysis. These results demonstrate that, even in complex clinical problems with limited understanding and predictive accuracy, our framework can still yield reliable outcomes. This is particularly significant, as most clinical prognostic models typically have low to moderate C-indices (0.5–0.7).

The second innovation of our framework addresses a major limitation in traditional trial emulation - unobserved confounding. As Hernán has noted, \say{target-trial emulation cannot overcome confounding bias from non-comparable treatment groups} \cite{hernan2021methods}. Our clinical problem - comparing chemotherapy with no adjuvant treatment - may be particularly vulnerable to unobserved confounding. As Hernán further emphasized: \say{confounding bias may be especially serious when emulating target trials that, [like ours], compare an active treatment with no treatment (or usual care) rather than with another active treatment.} \cite{hernan2021methods}. Since unobserved confounding, by definition, cannot be directly measured, how can we even know if it is present in this or any clinical problem? As Hernán pointed out, \say{it is generally impossible to determine whether the emulation failed because of uncontrolled confounding.} \cite{hernan2016using}. In our analysis, after matching for observed confounders and aligning baseline prognosis, we found that the predicted treatment benefit in our observational cohort was only 6\%, compared to 11\% in the target RCT. This discrepancy not only suggests the presence of unobserved confounding but also allows us to quantify its effect size. To mitigate this, we employed our weight-tuning method, which improved the predicted 5-year RFS estimates for the treated group by 5\%, bringing the observed treatment benefit to 11\%-matching the results of the JCOG0603 trial. This innovation demonstrates that our framework can account for unobserved confounding, a critical challenge in trial emulation.

After emulating the trial, we extended our analysis by identifying patient subgroups that may experience varying levels of benefit (or no benefit) from chemotherapy. While trials report the average treatment effect, this can obscure important heterogeneity in treatment effect-specifically, subsets of patients who do not benefit in an otherwise positive trial or who benefit in a trial with overall negative results. Using predicted 5-year RFS probabilities (for adjuvant chemotherapy or no chemotherapy) for each patient, we trained decision trees that grouped patients based on combinations of characteristics, allowing us to identify subgroups with differential treatment effects. Our approach goes beyond traditional one-variable-at-a-time subgroup analyses, which focus on single characteristics, such as age or tumor size. This traditional approach can be problematic because patients typically differ on multiple characteristics simultaneously, several of which may influence treatment benefit \cite{kent2018personalized}. Grouping patients based on only one characteristic that predicts treatment benefit overlooks the possibility that some patients within the group may not benefit from the treatment due to other characteristics. Moreover, the subgroups used in trial subanalyses are often determined manually, based on clinical expertise, and are frequently defined post-hoc or during the trial, which can compromise the validity of the subanalysis. In contrast, our framework offers an algorithmic approach to defining these subgroups through optimization.

Aside from the obvious benefits of precision medicine, there is another advantage in the use of OPTs. Thanks to OPTs, there might be no need to emulate hypothetical trials, as has been previously suggested. To elaborate, in our approach, we emulated an existing RCT comparing treatment versus placebo. Although trials like these are challenging to emulate due to the higher confounding in observational data comparing treatment vs. no treatment (compared to treatment A vs. B), the advantage of emulating such trials is that they are as \say{general} as possible for a given clinical problem. Once we have emulated these general trials, the OPTs can identify patient subgroups with varying levels of benefit (or no benefit), thus eliminating the need for more \say{specific} trials, which may be impractical or even unethical. For instance, after the initial general trial establishing imatinib as an effective post-surgery treatment for GIST, it would have been unethical to conduct a new RCT with a placebo arm to refine which patients benefit from imatinib \cite{bertsimas2024interpretable}. Instead, we can emulate the first general trial and use the resulting OPTs to identify those patients who do not benefit from imatinib. Similarly, in the case of the STRASS trial, which took a long time to recruit patients due to the rarity of RPS, it would have been impractical to conduct a new trial focused solely on histologic subtypes as potential predictors of radiation therapy benefit \cite{bonvalot2020preoperative}. However, since the STRASS trial was a general RCT with a placebo arm and an inclusive treatment arm, we could emulate it and use OPTs to define treatment benefit across specific histologic subtypes. In summary, by emulating a \say{general} RCT within our framework, the resulting OPTs can answer questions that would otherwise require additional, more specific trials. These trials, which are often impractical or even unethical, are typically the hypothetical trials that would otherwise need to be emulated. Since drug licensing systems internationally require RCT evidence of benefit or equivalence for nearly all new drugs, it is almost always possible to find an existing \say{general} trial to emulate, thus obviating the need for hypothetical trials.

So far, we have described a methodology that not only uses observational data to emulate a trial but also goes a step further by identifying patient subsets with benefits differing from the overall results of the trial. Achieving precision medicine using observational data that emulate a trial is a bold claim, and such a claim certainly requires validation. To ensure objective validation, we tested our decision trees using the gold standard - the trial we aimed to emulate. To avoid any biases, we did not have access to the trial data, and the validation was carried out by the research group that conducted the trial. Specifically, after completing our analysis, we provided only two OPTs that were trained using the constrained rewards estimation (see Section \ref{sec:opts}) and obtained the corresponding KM plots from the group that conducted the RCT, without refining the cohort or the OPTs after the validation.

The validation compared the actual effect of adjuvant chemotherapy - determined by KM plots - in the subsets defined by the OPT in the emulated cohort and the RCT cohort. The validation was fully successful: the subsets we identified in our emulated cohort that benefited from adjuvant chemotherapy also showed similar benefits in the trial cohort, with almost identical effect sizes. Likewise, the patient subsets in our emulated cohort that did not benefit from chemotherapy showed no benefit in the RCT cohort. Notably, the proportion of patients who did not benefit from adjuvant chemotherapy was strikingly consistent across both cohorts - 14\% in the emulated cohort and 15\% in the RCT cohort - further strengthening the credibility of our analysis. Thus, not only were we able to emulate a trial, but we also uncovered novel insights from the observational data that were not apparent in the trial itself. Specifically, we found that around 15\% of patients did not benefit from chemotherapy, a finding that was not evident in the original RCT analysis, which had indicated a benefit for all CRLM patients.

As mentioned earlier, CRLM is a particularly challenging problem due to the presence of unknown prognostic factors, likely genomic factors that we do not yet fully understand. One might wonder how our approach - or any approach - can fully address the clinical question of who benefits from chemotherapy, given the unknown factors that could be influencing treatment response and prognosis. To answer this, we examined a second OPT that split on different patient characteristics to identify subsets who did or did not benefit from chemotherapy. Importantly, although the characteristics used by the tree differed, the patients for whom both the new OPT and the previous OPT recommended chemotherapy were nearly identical, with a 95\% overlap in both cases. More importantly, the actual effect of adjuvant chemotherapy-determined by KM plots-in the subsets defined by the second OPT in both the emulated cohort and the RCT cohort was almost identical to that observed in the initial validation, and consistent between both cohorts. Notably, the proportion of patients who did not benefit from adjuvant chemotherapy in both the emulated and RCT cohorts was strikingly consistent across both OPTs-14\% vs 15\% for the first OPT and 16\% vs 17\% for the second OPT-further strengthening the credibility of our analysis. Taken together, these findings suggest that the combinations of patient characteristics defining subgroups with differing treatment benefits serve as stable proxies for clinical and potentially genomic factors, ensuring the reproducibility and replicability of the decision tree's recommendations.

In conclusion, we present a novel framework for emulating clinical trials using observational data, building on our earlier work in de-confounding such data. This framework was applied to solve a methodologically and clinically challenging problem and was successfully validated externally using an RCT cohort. Pending further validations by other researchers, this approach has the potential to profoundly impact both medical research and clinical practice in several key ways. First, by leveraging the abundance of real-world data, we can emulate RCTs with much larger sample sizes, addressing the low power issues that often encountered in traditional trials, which are constrained by the high cost and lengthy timelines required to recruit eligible patients. This is particularly important for negative trials, where our framework could help identify patient subsets who benefit from treatment, despite the overall negative results of the trial. Second, real-world data allows for the inclusion of variables not available in original trials-such as genomic data, which is increasingly accessible as technology advances-providing a richer and more nuanced understanding of treatment effects. Third, this approach supports faster and more efficient analysis by enabling multiple researchers, beyond just the group that owns the RCT data, to work on the emulated trial data. This collaborative access can accelerate the production of knowledge and foster the exploration of new ideas. Finally, it promotes greater inclusivity and fairness by allowing researchers with limited resources to recreate or emulate trials, democratizing access to high-quality data and contributing to more equitable scientific discovery.

\newpage

\bibliographystyle{apacite}
\bibliography{ref.bib}
\newpage

\setcounter{figure}{0}
\renewcommand{\thefigure}{S\arabic{figure}}
\FloatBarrier
\section{Supplementary Material}
\begin{figure}[hbtp]
\centering
\caption{Recurrence-free survival for patients defined by node 2 of the first OPT, in (A) the emulated cohort and (B) the RCT cohort}
\includegraphics[width=1\linewidth]{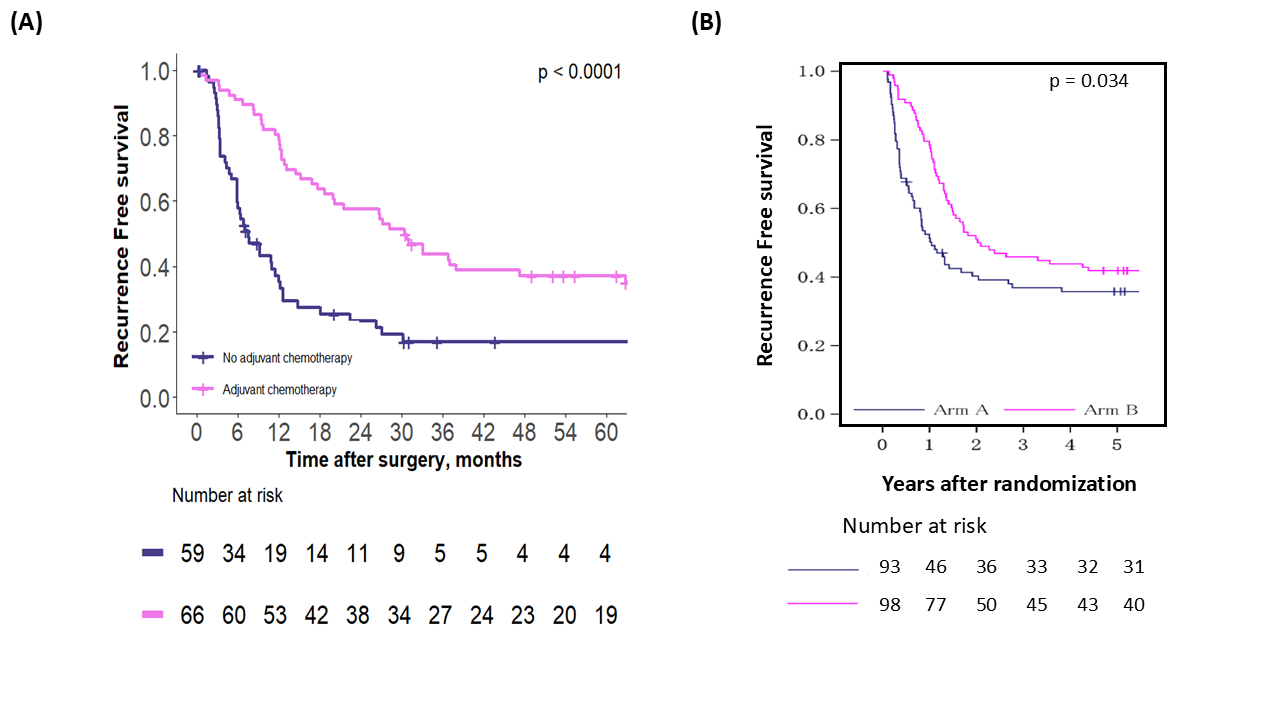}
\end{figure}

\begin{figure}[hbtp]
\centering
\caption{Recurrence-free survival for patients defined by node 5 of the first OPT, in (A) the emulated cohort and (B) the RCT cohort}
\includegraphics[width=1\linewidth]{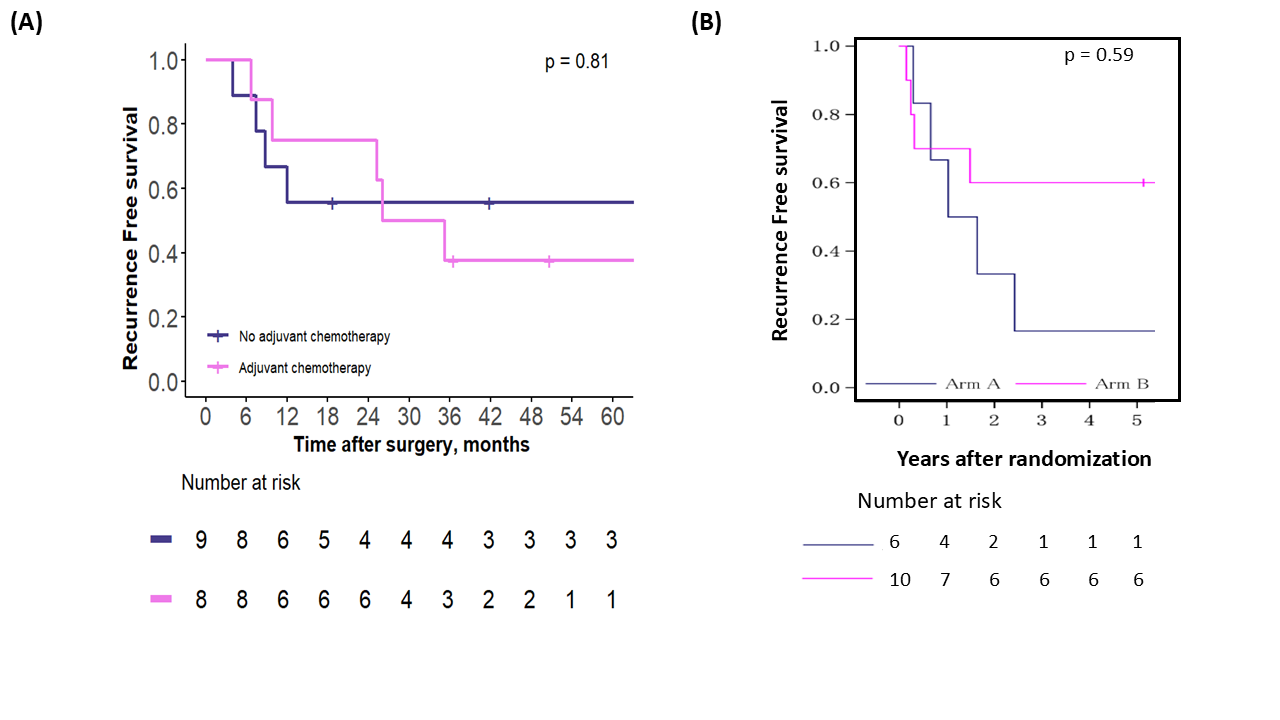}
\end{figure}

\begin{figure}[hbtp]
\centering
\caption{Recurrence-free survival for patients defined by node 7 of the first OPT, in (A) the emulated cohort and (B) the RCT cohort}
\includegraphics[width=1\linewidth]{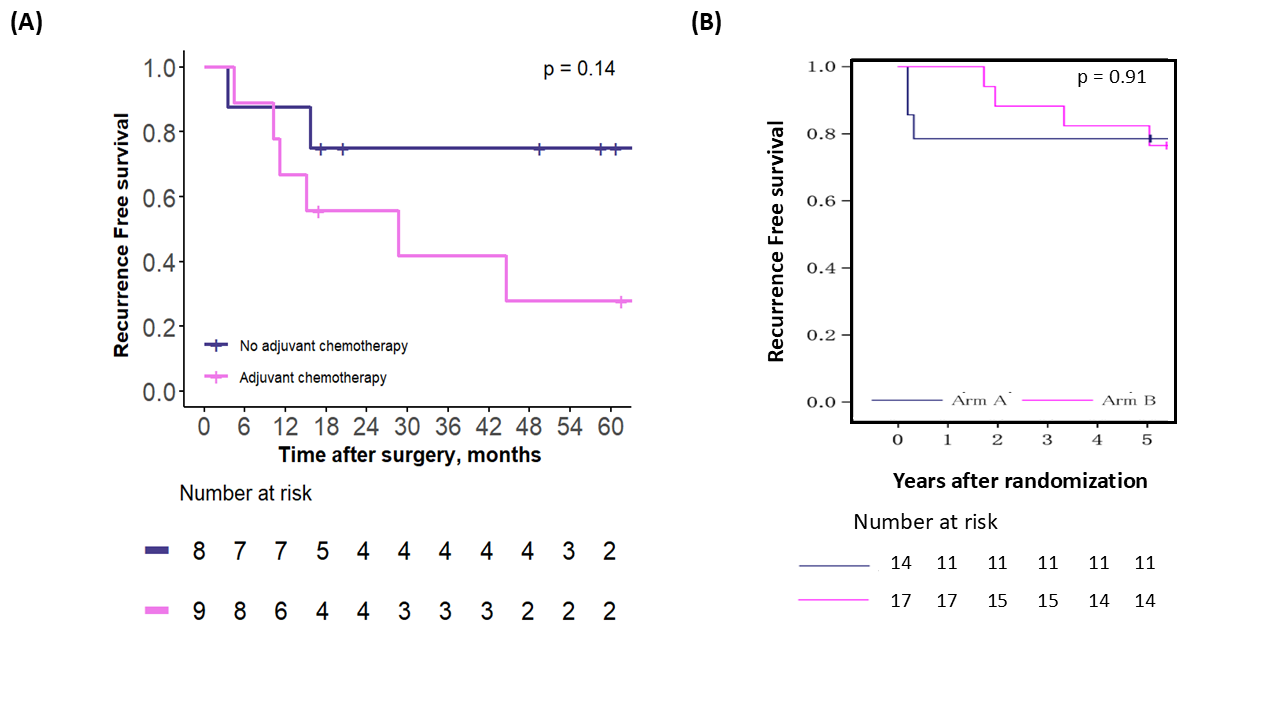}
\end{figure}

\begin{figure}[hbtp]
\centering
\caption{Recurrence-free survival for patients defined by node 8 of the first OPT, in (A) the emulated cohort and (B) the RCT cohort}
\includegraphics[width=1\linewidth]{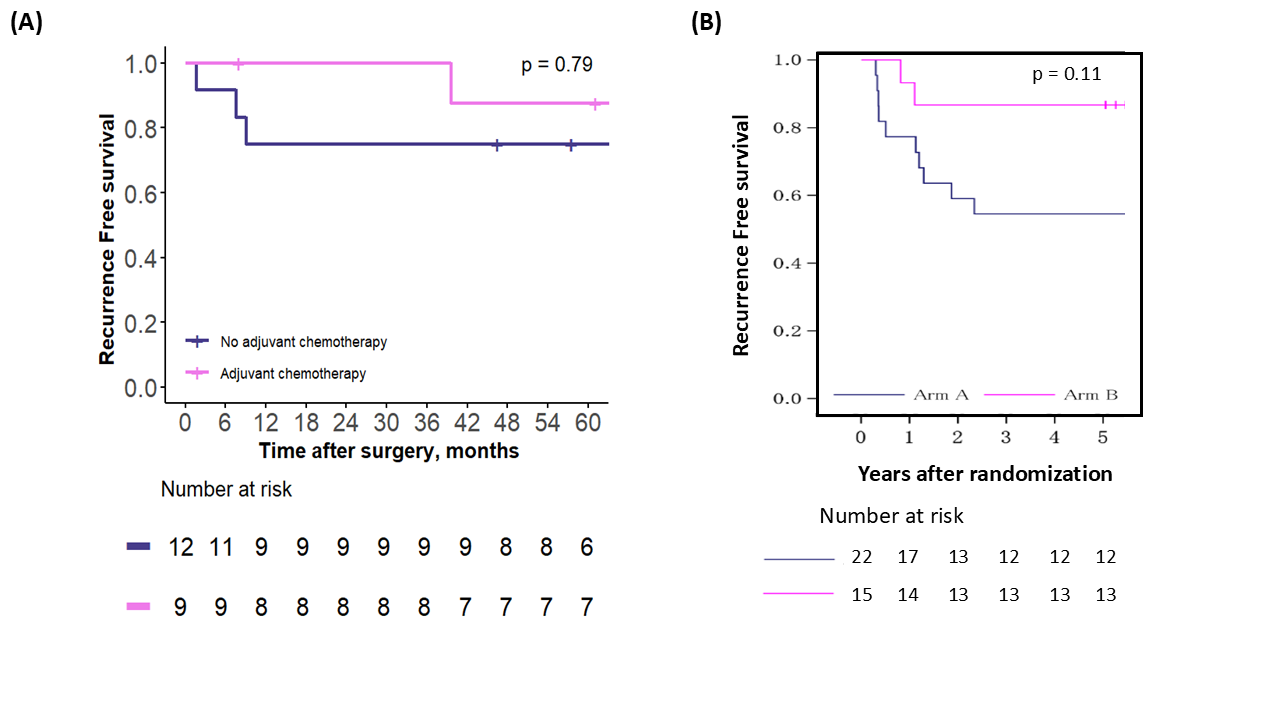}
\end{figure}

\begin{figure}[hbtp]
\centering
\caption{Recurrence-free survival for patients defined by node 9 of the first OPT, in (A) the emulated cohort and (B) the RCT cohort}
\includegraphics[width=1\linewidth]{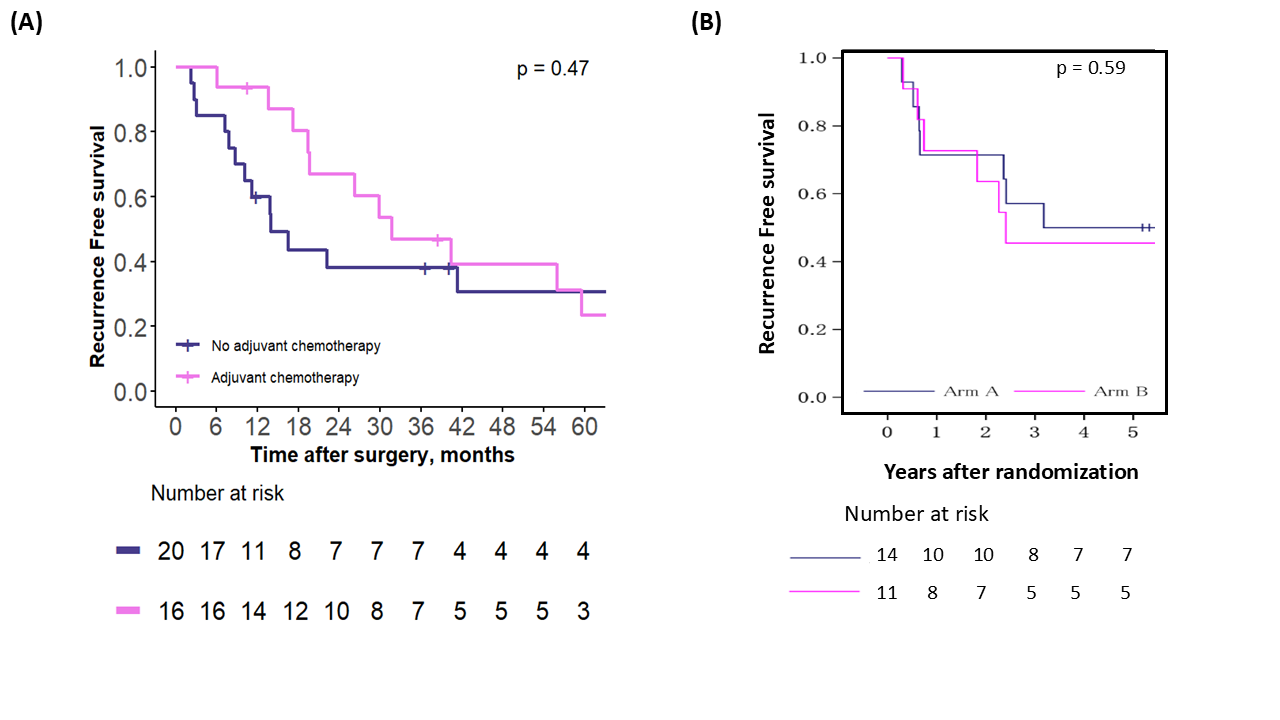}
\end{figure}

\begin{figure}[hbtp]
\centering
\caption{Recurrence-free survival for patients defined by node 2 of the second OPT, in (A) the emulated cohort and (B) the RCT cohort}
\includegraphics[width=1\linewidth]{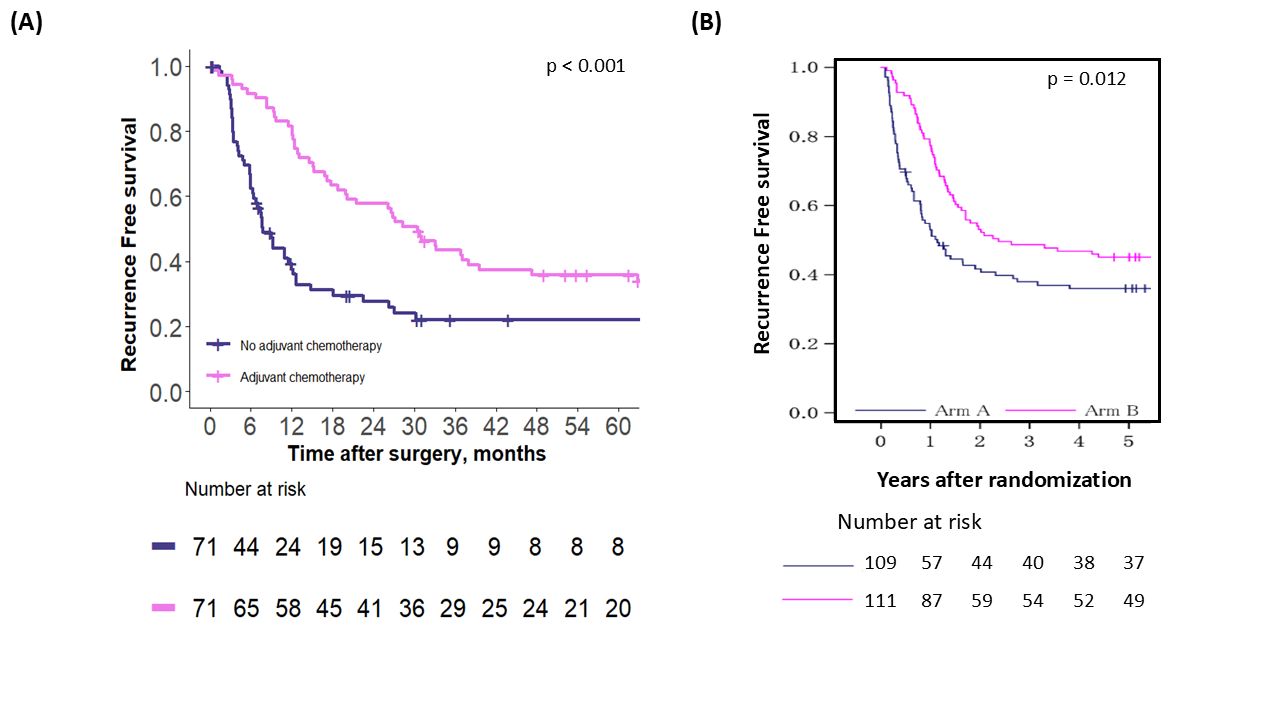}
\end{figure}

\begin{figure}[hbtp]
\centering
\caption{Recurrence-free survival for patients defined by node 5 of the second OPT, in (A) the emulated cohort and (B) the RCT cohort}
\includegraphics[width=1\linewidth]{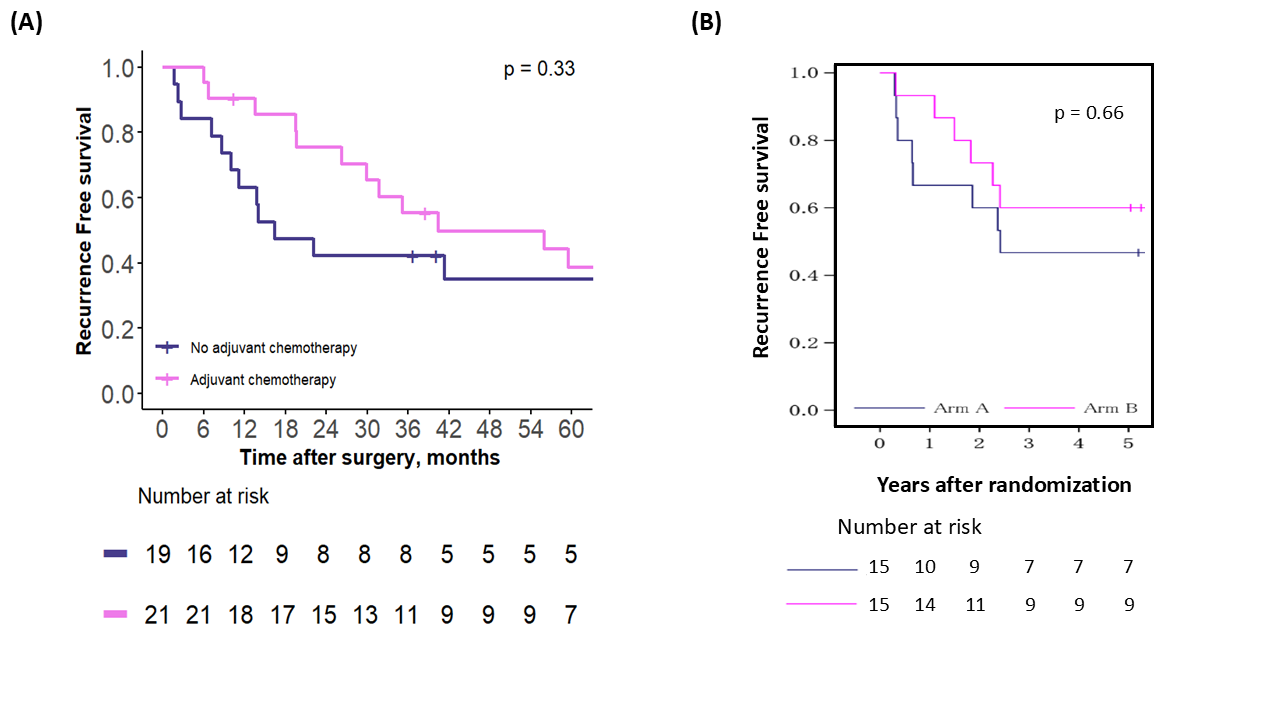}
\end{figure}

\end{document}